\documentclass[aps,superscriptaddress,preprintnumbers,amsmath,amssymb,prl,twocolumn]{revtex4-2}
\usepackage{tabularx}
\usepackage{graphicx}
\usepackage{amsmath}
\usepackage{color}
\usepackage{amssymb}
\usepackage{braket}
\usepackage[export]{adjustbox}
\usepackage{bm}
\usepackage{float}
\usepackage{array}
\usepackage{multirow}

\begin{document}

\title{Efficient First-Principles Framework for Overdamped Phonon Dynamics and Anharmonic Electron-Phonon Coupling in Superionic Materials}

\author{Yuxuan Wang}
\affiliation{Department of Materials Science and Engineering, University of Michigan, Ann Arbor, Michigan 48109, USA}
\author{Marios Zacharias}
\email{zachariasmarios@gmail.com}
\affiliation{Univ Rennes, INSA Rennes, CNRS, Institut FOTON - UMR 6082, F-35000 Rennes, France}
\affiliation{Computation-based Science and Technology Research Center, The Cyprus Institute, Aglantzia 2121, Nicosia, Cyprus}
\author{Xiao Zhang}
\affiliation{Department of Materials Science and Engineering, University of Michigan, Ann Arbor, Michigan 48109, USA}
\author{Nick Pant}
\affiliation{Department of Materials Science and Engineering, University of Michigan, Ann Arbor, Michigan 48109, USA}
\affiliation{Applied Physics Program, University of Michigan, Ann Arbor, Michigan 48109, USA}
\affiliation{Oden Institute for Computational Engineering and Sciences, The University of Texas at Austin, Austin, Texas 78712, USA}
\author{Jacky Even}
\affiliation{Univ Rennes, INSA Rennes, CNRS, Institut FOTON - UMR 6082, F-35000 Rennes, France}
\author{Pierre F. P. Poudeu} 
\affiliation{Department of Materials Science and Engineering, University of Michigan, Ann Arbor, Michigan 48109, USA}
\author{Emmanouil Kioupakis}
\email{kioup@umich.edu}
\affiliation{Department of Materials Science and Engineering, University of Michigan, Ann Arbor, Michigan 48109, USA}

\date{\today}

\begin{abstract}
 Relying on the anharmonic special displacement method, we introduce an {\it ab initio} quasistatic polymorphous framework to describe local disorder, anharmonicity, and electron-phonon coupling in superionic conductors. Using cubic Cu$_2$Se, we show that positional polymorphism yields the breakdown of the phonon quasiparticle picture leading to extremely overdamped anharmonic vibrations while preserving transverse acoustic phonons, consistent with experiments. 
 We also demonstrate highly broadened electronic spectral functions with band gap openings of 1.0~eV due to polymorphism, and that anharmonic electron-phonon coupling leads to a band gap narrowing with increasing temperature. Our approach, relying on generating a handful of configurations, opens the way for efficient calculations in superionic crystals to elucidate their compelling high figure-of-merit.
\end{abstract}

\maketitle

Copper chalcogenides Cu$_2$(Se, S, Te) and related compounds have recently attracted significant attention for their unique properties and potential in thermoelectric applications~\cite{padamPropertiesChemicallyDeposited1987,liuCopperIonLiquidlike2012,liuEnhancingThermoelectricPerformance2015, yangHighperformanceThermoelectricCu2Se2015,liu_reduction_2016,Bailey2016,voneshenHoppingTimeScales2017,Olvera2017,Zhang2020_review,Hu2024_NatMater,Zhang2024_NatCommun}, achieving a remarkable figure-of-merit ($zT$) above 2, compared to typical values around 1~\cite{liuUnderstandingContactNanostructured2013, yanExperimentalStudiesAnisotropic2010}. The high-temperature phase of these materials is characterized by a cubic superionic structure, featuring a disordered arrangement of Cu cations~\cite{Zhang2020_review,heydingCrystalStructuresCu11976,stevelsTransitionT,TransitionT_1945,Takahashi1976,Evans1979,Asadov1992,Gulay2011,Roth2019,Kavirajan2021,Jain2023}. The specific transition temperature is influenced by the degree of Cu deficiency, $x$, which can stabilize the cubic phase of Cu$_{2-x}$Se even at room temperature~\cite{heydingCrystalStructuresCu11976,stevelsTransitionT,TransitionT_1945}. 

It has been proposed that, similar to liquids, superionic diffusion of Cu ions in Cu$_2$Se suppresses phonon modes, while the electron behavior remains consistent with crystalline semiconductors~\cite{liuCopperIonLiquidlike2012}. This Phonon-Liquid Electron-Crystal (PLEC) phenomenon has been challenged in Ref.~[\onlinecite{voneshenHoppingTimeScales2017}], suggesting that Cu hopping is too slow to suppress transverse acoustic phonons. The high $zT$ has been attributed to anharmonic lattice dynamics rather than liquid-like diffusion, highlighting the need for a deeper understanding of the interplay between ionic disorder and lattice vibrations in superionic conductors.

Molecular dynamics (MD) simulations~\cite{kimUltralowThermalConductivity2015,namsaniThermalConductivityThermoelectric2017,zhuo_liquidlike_2020} have made excellent progress in describing vibrational properties, including the low thermal conductivity. However, these simulations do not effectively distinguish between the contributions of atomic disorder and thermal disorder to the overall vibrational behavior, thus limiting their ability to fully elucidate the mechanisms underlying the high $zT$. Alternative approaches to vibrational dynamics in superionic conductors~\cite{zhangAnharmonicPhononFrequency2020} involve phonon anharmonicity calculations~\cite{tadanoSelfconsistentPhononCalculations2015} based
on the self-consistent phonon (SCP) theory~\cite{hootonLINewTreatment1955}.
While these calculations capture phonon self-energy corrections at finite temperatures~\cite{tadanoSelfconsistentPhononCalculations2015,Hellman2013,Errea2014,zachariasAnharmonicLatticeDynamics2023}, they rely on the phonon quasi-particle approximation and do not account for atomic disorder. This results in substantial discrepancies with experimental measurements of the phonon density of states (PDOS)~\cite{liu_reduction_2016,voneshenHoppingTimeScales2017}.

The positions of Cu atoms also impact the electronic properties, leading to, for example,  significant band gap changes. Semilocal density functional theory (DFT) calculations~\cite{LDA,perdewGeneralizedGradientApproximation1996} predict a semi-metallic behavior in Cu$_2$Se when using a simulation cell with atoms at their high-symmetry positions~\cite{rasanderDensityFunctionalTheory2013,tyagiBandStructureTransport2014,zhangElectronicStructureAntifluorite2014}.
 Improved treatments of exchange or correlation using hybrid~\cite{krukauInfluenceExchangeScreening2006,pbe0} or the modified Becke-Johnson~\cite{PhysRevLett.102.226401,judgemBJ} functionals open a band gap of 0.5 eV~\cite{zhangElectronicStructureAntifluorite2014,rasanderDensityFunctionalTheory2013,sunElectronCrystalBehavior2017},  underestimating experiments ranging from 0.8 to 1.5~eV~\cite{padamPropertiesChemicallyDeposited1987,Gad2022,soviet_gap,Marshall1965}. Using these functionals, empirical adjustments to the atomic positions and MD simulations for modeling atomic disorder can increase the gap up to 1.0~eV~\cite{lukashev,zhangElectronicStructureAntifluorite2014,sunElectronCrystalBehavior2017,klanTheoreticalStudyImpact2021}. 

\begin{figure*}[!tbp]
  \centering
\includegraphics[width=0.95\textwidth]{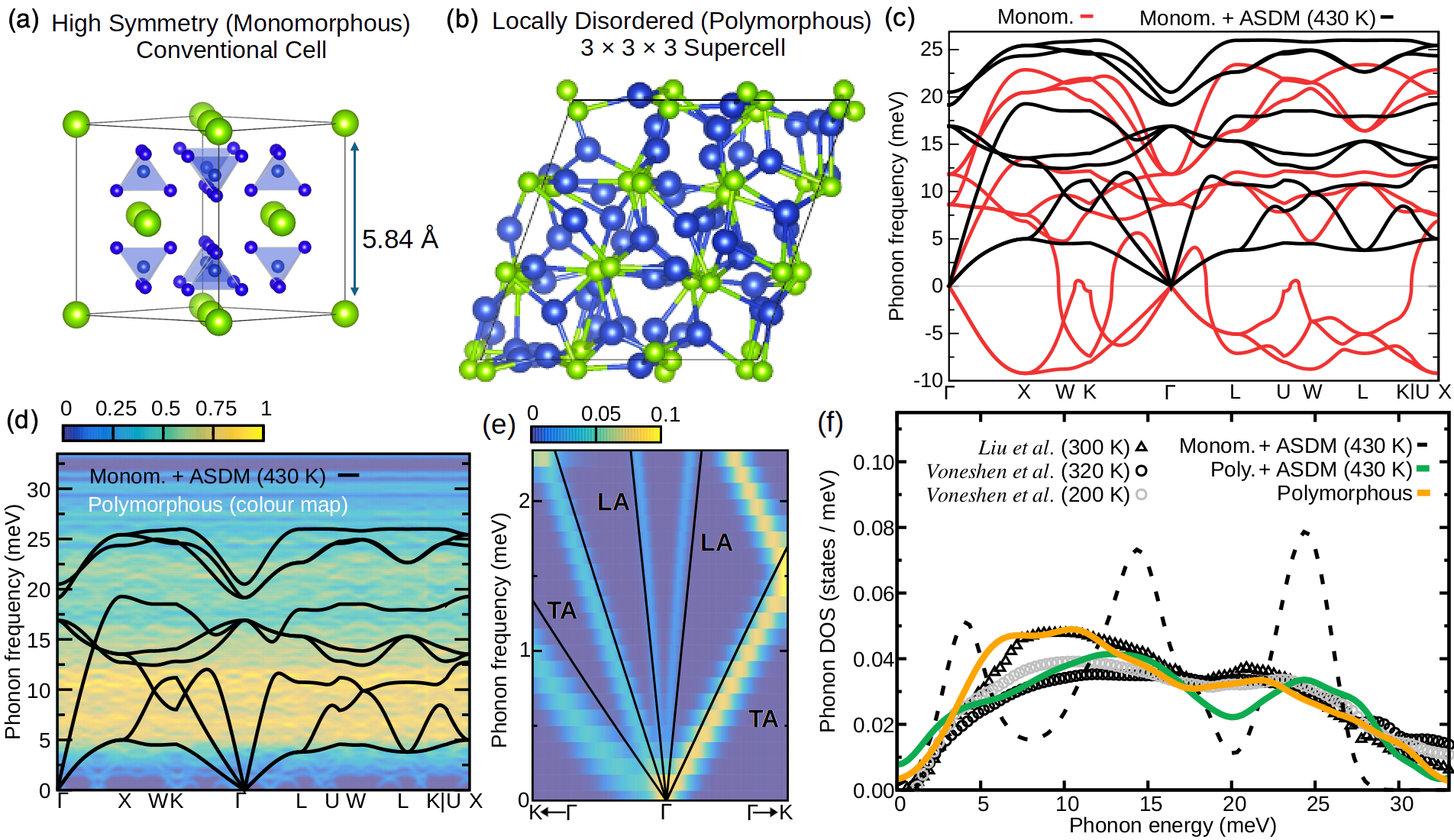}
 \caption{\label{fig1}  (a) Conventional cell of cubic Cu$_2$Se~\cite{momma_vesta_2011}, showing the high-symmetry structure (Fm-3m) with four equivalent Cu 32f sites forming a tetrahedron centered at the symmetric 8c Wyckoff position\cite{heydingCrystalStructuresCu11976,danilkinNeutronDiffractionStudy2011,danilkinCrystalStructureLattice2003,SKOMOROKHOV200664}. The lattice parameter is $5.84$~\AA ~{\cite{danilkinNeutronDiffractionStudy2011}}.  (b) Polymorphous structure of cubic Cu$_2$Se in a $3\times 3\times3$ supercell of the primitive cell. (c)Harmonic phonon dispersion of cubic Cu$_2$Se for the monomorphous structure (red) and anharmonic dispersion at 430 K (black), calculated via the ASDM in $3\times3\times3$ supercells using the monomorphous structure to symmetrize the IFCs. (d) Phonon spectral function (color map) calculated for the polymorphous structure using phonon unfolding~\cite{zachariasAnharmonicElectronphononCoupling2023} for 1353 equally spaced {\bf q}-points. The black curve represents the ASDM dispersion at 430 K as in (c). (e) As in (d), but focused around the $\Gamma$-point, highlighting the transverse acoustic (TA) and longitudinal acoustic (LA) modes. (f) Phonon density of states computed using the monomorphous structure with ASDM at 430K (dashed black), the polymorphous structure (orange),  the polymorphous structure with ASDM at 430 K (green), all evaluated with 1353 {\bf q}-points. Experimental data (black and grey) at 300, 320, and 200 K are from Refs.~[\onlinecite{liu_reduction_2016}]  and~[\onlinecite{voneshenHoppingTimeScales2017}]. The calculated PDOS and data from Ref.~[\onlinecite{voneshenHoppingTimeScales2017}] are scaled by 8 and 4 to match data from Ref.~[\onlinecite{liu_reduction_2016}].}
 \end{figure*}

In this paper, motivated by the progress of previous studies~\cite{kimUltralowThermalConductivity2015,namsaniThermalConductivityThermoelectric2017,zhuo_liquidlike_2020,zhangAnharmonicPhononFrequency2020,rasanderDensityFunctionalTheory2013,tyagiBandStructureTransport2014,zhangElectronicStructureAntifluorite2014,sunElectronCrystalBehavior2017,klanTheoreticalStudyImpact2021,lukashev}, we introduce a systematic method to study disorder and anharmonicity, separating their impact on phonon dynamics and electronic structure.
We propose using a locally disordered structure, namely the polymorphous network~\cite{zhaoPolymorphousNatureCubic2020, Zhao2024}, whose atoms are fixed in positions that define a minimum on the potential energy surface in  the cubic phase. As previously shown for other strongly anharmonic materials~\cite{trimarchiPolymorphousBandStructure2018, zhaoPolymorphousNatureCubic2020,Goesten2022,zachariasAnharmonicElectronphononCoupling2023,Zhao2024}, polymorphous networks in cubic phases yield significant corrections to the electronic structure and phonons compared to calculations based on the high-symmetry (monomorphous) structure.

Here, we employ the Anharmonic Special Displacement Method (ASDM)~\cite{zachariasAnharmonicElectronphononCoupling2023, zachariasAnharmonicLatticeDynamics2023, zachariasTheorySpecialDisplacement2020} to explore polymorphous structures, anharmonicity, and electron-phonon coupling in cubic Cu$_2$Se. Our calculations show that accounting for local disorder yields overdamped vibrations across the entire phonon spectrum and reproduces precisely previous measurements of the PDOS~\cite{liu_reduction_2016,voneshenHoppingTimeScales2017}, without suppressing transverse acoustic phonons.  Furthermore, our static DFT calculations for the polymorphous network  reveal the breakdown of the electron quasiparticle approximation. HSE06 and PBE0 calculations~\cite{krukauInfluenceExchangeScreening2006,pbe0} yield a band gap of 0.89 and 1.45 eV, respectively, a significant improvement compared to previous calculations using monomorphous structures or  semi-empirical approaches. Our anharmonic electron-phonon coupling calculations using the polymorphous structure yield a band gap reduction due to the quantum zero-point motion of 35~meV and a further temperature-induced band gap reduction of 212~meV from 0 to 780 K. We also employ the Burstein-Moss shift~\cite{bursteinAnomalousOpticalAbsorption1954} to account for the heavy p-type doping due to Cu deficiency, and we obtain a band gap widening of 0.26 eV. Taken all contributions together, our values range from 0.8 to 1.7 eV at 300 K, matching experiment well~\cite{padamPropertiesChemicallyDeposited1987,Gad2022,soviet_gap,Marshall1965}.  Our approach is efficient, as it leverages the strengths of the ASDM method: it avoids long equilibration times to capture local disorder, requires only a handful of configurations to explore anharmonicity, and needs just a pair of configurations to determine the band gap at each temperature.


Our atomistic calculations are based on DFT, with a combination of semilocal and hybrid functionals. Semilocal DFT calculations are performed in {\tt QUANTUM ESPRESSO} (QE)~\cite{giannozziAdvancedCapabilitiesMaterials2017, giannozziQUANTUMESPRESSOModular2009} using plane wave basis sets and a cutoff of 100 Ry. We employ the Perdew-Burke-Ernzerhof exchange-correlation functional~\cite{perdewGeneralizedGradientApproximation1996} and Optimized Norm-Conserving Vanderbilt pseudopotentials~\cite{PhysRevB.88.085117}. To obtain polymorphous structures, we employ the ASDM as implemented in the {\tt ZG} package of the {\tt EPW} code~\cite{leeElectronphononPhysicsFirst2023} and the strategy in Refs.~\cite{zachariasAnharmonicLatticeDynamics2023,zachariasAnharmonicElectronphononCoupling2023}. That is, we apply special displacements on the nuclei of the high-symmetry Cu$_2$Se [e.g. Fig.~\ref{fig1} (a)] along harmonic soft phonons in a supercell and relaxed the system while keeping the lattice constant fixed. This allows the generation of polymorphous structures [Fig.~\ref{fig1}(b)], yielding an energy lowering of up to 175~meV per formula unit compared to the monomorphous  primitive cell (3 atoms).  We emphasize that this procedure yields a locally disordered structure consistent with the cubic phase, and not a transition toward a low-temperature phase. A similar result and energy lowering were obtained using stable harmonic phonons, instead of soft modes. The majority of the calculations are performed on $3\times3\times3$ supercells  of the primitive cell using a uniform $3\times3\times3$ {\bf k}-grid, unless otherwise stated. HSE06 and PBE0 calculations are performed using VASP~\cite{blochlProjectorAugmentedwaveMethod1994, kresseEfficiencyAbinitioTotal1996}, with a cutoff of 370 eV and the relaxed structures obtained from QE. 
To evaluate electron and phonon spectral functions of the polymorphous networks, we use electron and phonon band structure unfolding~\cite{Allen2013} as implemented in {\tt ZG.x}~\cite{zachariasTheorySpecialDisplacement2020, zachariasAnharmonicElectronphononCoupling2023}. To account for anharmonic electron-phonon coupling in monomorphous and polymorphous structures, we apply special displacements on the nuclei using temperature-dependent anharmonic phonons computed by the ASDM. Full computational details are available in the supplemental information (SI)~\cite{si}.

\begin{figure*}[!tbp]
    \centering  \includegraphics[width=.98\textwidth]{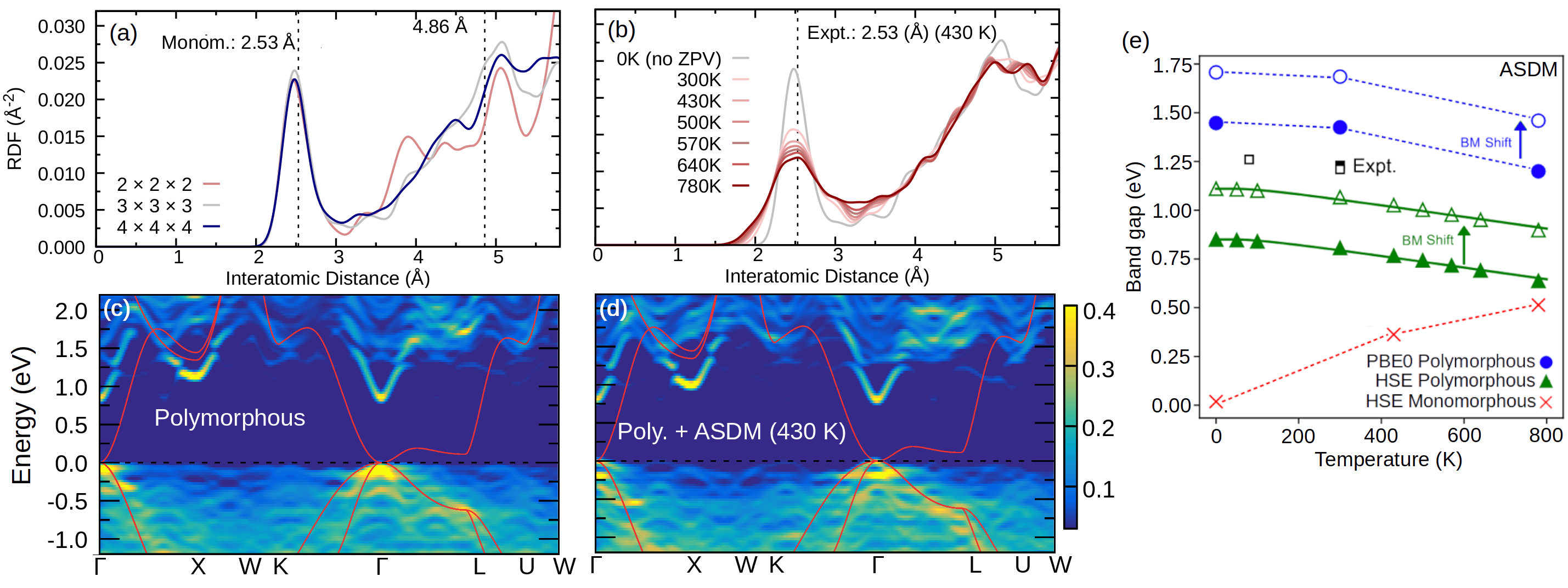}
  \caption{\label{fig2} (a,b) Cu-Se bonding Radial Distribution Function (RDF) calculated with MAISE~\cite{ibarra-hernandezStructuralSearchStable2018} for (a) $2\times2\times2$, $3\times3\times3$, and $4\times4\times4$ polymorphous supercells at 0~K without zero-point vibrations (ZPV) and (b) $3 \times 3 \times 3$ supercells using the ASDM on top of the polymorphous network to account for thermal disorder, from 300 to 780 K. The experimental value is from Ref.~[\onlinecite{danilkinNeutronDiffractionStudy2011}]. 
   (c,d) DFT electron spectral functions calculated using $3\times3\times3$ polymorphous supercells and  band structure unfolding~\cite{zachariasTheorySpecialDisplacement2020} without (c) and with anharmonic electron-phonon renormalization at 430~K (d). The conduction band energies are shifted by 0.7~eV to match the HSE band gap. The band structure calculated for the monomorphous cell is shown in red.
  (e) HSE (green) and PBE0 (blue) temperature-dependent band gaps calculated via ASDM for $3 \times 3 \times 3$ polymorphous supercells. The corresponding HSE data for the monomorphous structure are shown in red. Green solid lines are single Einstein oscillator fits to the HSE gaps of the polymorphous structure. Red and blue dashed lines are used as a guide to the eye. Empty circles and triangles represent Burstein-Moss (BM) shifted data by 0.26 eV. All ASDM data represent the average band gap calculated for an antithetic pair~\cite{Zacharias2016}. Experimental values of Cu$_{1.9}$Se (filled square) and Cu$_{2}$S (empty squares) are from Refs.~\cite{padamPropertiesChemicallyDeposited1987} and~\cite{Marshall1965}.}
\end{figure*}

 Figure ~\ref{fig1}(c) shows the phonon dispersions of cubic Cu$_2$Se obtained within the phonon quasiparticle picture using interatomic force constants calculated for the monomorphous structure with and without anharmonic corrections. The harmonic phonon dispersion calculated using the monomorphous structure exhibits large instabilities across the entire reciprocal space, which are eliminated by accounting for phonon self-energy corrections at 430~K within the ASDM. In the SI, we provide the temperature dependence of the phonon dispersion [Fig.~S1(a)] and the temperature dependence of the optical phonons at the zone-center, accounting for the longitudinal-transverse optical mode splitting due to long-range effects [Table I and Figs.~S1(b), S2~\cite{si}]. The ASDM convergence performance for evaluating anharmonic phonons at 430~K is shown in Figs. S3 and S4~\cite{si}.  In Fig. S5, we verify that instabilities persist in the harmonic approximation even when using PBE0 and HSE or different pseudopotentials.

We then examine the effects of atomic disorder on the vibrational dynamics. In Fig.~\ref{fig1}(d), we present the phonon spectral function of polymorphous Cu$_2$Se obtained by phonon unfolding. Our results reveal strongly overdamped vibrations, deviating from the quasi-particle band theory [monomorphous + ASDM in Fig.~\ref{fig1}(d)]. The slow ionic diffusion~\cite{voneshenHoppingTimeScales2017}, mimicked by our quasistatic polymorphous network, leads to strongly correlated vibrations with finite lifetimes, suggesting significant phonon-phonon scattering rates that limit thermal conductivity. Examining the low-frequency region around the zone-center [Fig.~\ref{fig1}(e)], we find that local disorder preserves transverse acoustic modes, inconsistent with a phonon-liquid picture. To validate the use of the polymorphous structure for describing vibrational properties, we compute the PDOS and compare it with measurements~\cite{liu_reduction_2016,voneshenHoppingTimeScales2017}, as shown in Fig.~\ref{fig1}(f). The PDOS computed for the polymorphous network yields excellent agreement with experiments across the phonon energy range of 0 - 35~meV, exhibiting large broadening without sharp features. In contrast, standard anharmonic approaches using monomorphous cells and accounting only for temperature-dependent self-energy corrections  fail to reproduce experiment, resulting in a narrower frequency range of 0 - 25~meV and sharp peaks [monomorphous + ASDM in Fig.~\ref{fig1}(f)].  A similar result using two different implementations of the SCP theory~\cite{Hellman2013,tadanoSelfconsistentPhononCalculations2015} for the monomorphous Cu$_2$Se has been reported in Ref.~[\onlinecite{zhangAnharmonicPhononFrequency2020}]. Notably, experiments show that the PDOS remains almost unchanged over the temperature range 320 -- 900 K~\cite{voneshenHoppingTimeScales2017}.

Overall, our results in Fig.~\ref{fig1}(f) suggest  the breakdown of the phonon quasiparticle picture where local disorder plays a more significant role in capturing anharmonicity in superionic systems than self-energy corrections in monomorphous models derived from higher-order diagrams~\cite{tadanoSelfconsistentPhononCalculations2015,Bianco2017,zachariasAnharmonicLatticeDynamics2023}.  This is further demonstrated in supplemental Fig. S6, which shows the phonon spectral function computed by combining local disorder with self-energy corrections, maintaining the vibrational features. 
 Moreover, self-energy corrections at 430~K combined with local disorder [green in Fig.~\ref{fig1}(f)] preserve the broad features of the PDOS and do not introduce sharp peaks. This confirms that their overall impact is relatively small compared to the failure of using self-energy corrections alone within the quasiparticle picture based on a monomorphous structure.
The success of using a quasistatic polymorphous network, with atoms localized in one of the potential energy wells, aligns with slow, low-energy hopping relaxations. These localized dynamics minimally disrupt low-energy vibrations, allowing for transverse acoustic phonon propagation despite the extreme anharmonicity, similar to findings for ultrasoft halide perovskites~\cite{Ferreira2020,zachariasAnharmonicElectronphononCoupling2023}.

 We next analyze the effects of disorder and phonon anharmonicity on the radial distribution functions (RDFs) [Eq.~(1) in SI~\cite{si}]. In Fig.~\ref{fig2}(a), we report the RDF 
of Cu-Se bonds calculated using polymorphous structures generated for different supercell sizes without considering thermal effects. In Fig. S7~\cite{si}, we also report Cu-Cu and Se-Se RDFs which compare well with experiments~\cite{danilkinNeutronDiffractionStudy2011} and MD calculations~\cite{zhangAnharmonicPhononFrequency2020}. Unlike the RDF of the monomorphous network, represented by the vertical dashed lines in Fig.~\ref{fig2}(a), the RDFs for the polymorphous structures display broad features and indicate a wide distribution of bond lengths  beyond the first peak, reflecting local structure within the system.
Figure~\ref{fig2}(a) and data in supplemental Table II~\cite{si} show small differences between the RDFs calculated for different supercell sizes, suggesting that bond length statistics are captured well even by the $2\times2\times2$ supercell.
We also observe that the shortest Cu-Se bond peak position deviates slightly (1.9 \% away) from the monomorphous cell value and the experimental value of 2.53 \AA\, at 430 K~\cite{danilkinNeutronDiffractionStudy2011}. Figure ~\ref{fig2}(b) shows the RDF in the temperature range 300 -- 780~K calculated using $3\times3\times3$ supercells that account for both local and thermal disorder via the ASDM. Anharmonic vibrational dynamics lead to the shift of the first RDF peak, improving agreement between theory and experiment. This highlights the importance of both local and thermal disorder in reproducing structural features. The thermal disorder also leads to a larger broadening and decrease in amplitude of the first RDF peak. For example, the RDF peak calculated for $430$~K has a full width at half maximum (FWHM) of 0.7 \AA\, compared to 0.32~\AA\ when no thermal disorder is introduced. The temperature dependence of the first RDF peak position and FWHM are available in the SI Table III~\cite{si}. 
 


Subsequently, we determine the effects of structural and thermal disorder on the electronic properties.  In Figs.~\ref{fig2}(c,d), we report our calculations for the electron spectral function of  polymorphous cubic Cu$_2$Se using the static configuration and the configuration generated for 430~K by the ASDM. Both spectral functions are calculated for $3\times3\times3$ supercells using  band unfolding~\cite{zachariasTheorySpecialDisplacement2020} and 382 {\bf k}-points. The monomorphous structure, shown as red, yields a semi-metallic behavior  even at the HSE level (Fig.~S8), consistent with previous results~\cite{rasanderDensityFunctionalTheory2013,tyagiBandStructureTransport2014,zhangElectronicStructureAntifluorite2014}.  Importantly, positional polymorphism induces significant broadening, to the point where the valence band states are significantly smeared out, confirming that the electron quasiparticle approximation breaks down in this regime, as observed in other materials~\cite{LihmArxiv2025}. Furthermore, local disorder opens an HSE gap at the $\Gamma$-point of 0.88 eV. This value converges within 0.1 eV with our HSE band gaps of
 0.82 and 0.89 eV obtained for  $2\times2\times2$ and $4\times4\times4$ supercells, respectively (Fig. S9~\cite{si}). Figure~\ref{fig2}(d), shows that electron-phonon coupling at 430~K, within the adiabatic approximation, leads to a small redistribution of the spectral weights and a band gap closing of approximately 0.1 eV.  Unlike the valence band, dispersive bands close to the conduction band edges are preserved, suggesting that the electron effective masses remain optimal for charge transport and thermoelectric performance.

Figure~\ref{fig2}(e) shows our HSE calculations for the temperature-dependent band gap of monomorphous and polymorphous cubic Cu$_2$Se, accounting for anharmonic electron-phonon coupling within the ASDM. Our calculations for the monomorphous structure show a band gap widening as the temperature increases, contrary to experimental measurements in Cu-based compounds~\cite{Marshall1965,Choi2014,Birkett2018,qasrawi_situ_2019,iraq_study_2022,henry_temperature_2021}. 
The decrease in the band gap with temperature, typical for traditional semiconductors~\cite{Allen1976,Allen1981,Giustino2010,Ponc2015,Zacharias2016,Miglio2020,zachariasTheorySpecialDisplacement2020,Engel2022}, is captured correctly when we account for positional polymorphism [Fig.~\ref{fig2}(e)], giving a monotonic band gap decrease of 0.04 eV from 50 to 300 K. This agreement highlights the significant influence of local disorder on electron-phonon coupling in Cu-based superionic conductors, as previously observed in polymorphous perovskites~\cite{zachariasAnharmonicElectronphononCoupling2023}.  We further apply a Bose-Einstein oscillator fit~\cite{Giustino2010} of the form $-\alpha-2\alpha/(e^{\Theta/T} - 1)$ (green line), and find $\alpha$ = 35~meV and $\Theta=269$~K, which represent the zero-point renormalization and the effective phonon temperature, respectively. From $\Theta$, we extract an effective energy of 23~meV, demonstrating that high-energy optical vibrations dominate band gap renormalization in Cu$_{2}$Se. We emphasize that the supercell size is critical for describing the electronic structure and electron-phonon coupling, as shown in Figs. S9, S10, and S11~\cite{si}. Our data is also well described by the Varshni relation~\cite{si,varshniTemperatureDependenceEnergy1967} [Fig. S12]. We note that our ASDM calculations are based on the adiabatic Allen-Heine theory~\cite{Allen1976,Allen1981,Ponc2014} and thus neglect non-adiabatic effects, which can be important for phonon-induced band gap renormalization in some materials~\cite{Ponc2015,Miglio2020,Engel2022}. Given the relatively low phonon energies in Cu$_2$Se, we expect non-adiabatic effects to introduce quantitative refinements but not alter our conclusions.

Despite accounting for local disorder and electron-phonon coupling, our HSE band gap at 300 K still underestimates experiment~\cite{padamPropertiesChemicallyDeposited1987} by 0.33 eV. To account for the heavy p-type doping due to 5\% Cu deficiency in the experimental samples~\cite{Zhang2020_review}, we also evaluate the Burstein-Moss (BM) shift~\cite{bursteinAnomalousOpticalAbsorption1954,si} due to band filling and Pauli blocking. We assume Cu$_{1.9}$Se, a three-fold degenerate parabolic band edge with spin degeneracy, and use the experimental effective mass~\cite{voskanyana.a.1978,si}. This results in an additional 0.26 eV gap opening for Cu$_{1.9}$Se [empty triangles in Fig.~\ref{fig2} (e)], further improving the agreement with the experimental value of 1.23 eV~\cite{padamPropertiesChemicallyDeposited1987}. For completeness, in Fig.~\ref{fig2}(e), we also report the electron-phonon renormalized band gap at the PBE0 level, which is 1.45 eV at 0 K and 1.43 eV at 300 K. Combined with a BM shift of 0.26 eV (empty circles), these values establish an upper bound, consistent with experimental results at room temperature that span a wide range~\cite{padamPropertiesChemicallyDeposited1987,Gad2022,soviet_gap,qasrawi_situ_2019,iraq_study_2022,henry_temperature_2021}.

In conclusion, we introduce an efficient  and broadly applicable  first-principles framework to simulate strong anharmonicity and electron-phonon coupling in superionic conductors without MD simulations. Relying on the ASDM, we demonstrate that quasistatic polymorphous networks of cubic Cu$_2$Se lead to (i) strongly overdamped vibrational dynamics without suppressing transverse acoustic modes; (ii) an accurate description of the PDOS, supporting the concept that atoms exhibit anharmonic vibrations and disorder, rather than translational diffusion characteristic of a liquid,  in contrast to the PLEC model; (iii)  significantly broad electron spectral functions; (iv) a large band gap opening compared to the monomorphous structure; and (v) a correct treatment to electron-phonon coupling compared to monomorphous structures, resulting in a decreasing band gap with increasing temperature, consistent with experiments. The present work lays the foundations for systematic first-principles calculations of the thermal conductivity, heat capacity, electrical conductivity, and Seebeck coefficients at finite temperatures, determining the performance of superionic thermoelectrics.

 Data supporting this study (inputs, structures, and outputs) are available via the NOMAD repository~\cite{NOMADdoi}.

This work is supported by National Science Foundation Award No. 2114424. Computational resources are from National Energy Research Scientific Computing (NERSC) Center (Contract No. DEAC02-05CH11231) for phonons and Advanced Cyber infrastructure Coordination Ecosystem: Services and Support (ACCESS) award DMR200031 for band structure calculations. M.Z. acknowledges funding by the European Union (project ULTRA-2DPK / HORIZON-MSCA-2022-PF-01 / Grant Agreement No. 101106654). Views and opinions expressed are however those of the authors only and do not necessarily
reflect those of the European Union or the European Commission. Neither the European Union nor the
granting authority can be held responsible for them. We also acknowledge computational resources from the EuroHPC Joint Undertaking and supercomputer LUMI [https://lumi-supercomputer.eu/], hosted by CSC (Finland) and the LUMI consortium through a EuroHPC Extreme Scale Access call.


\bibliography{ref-final}

\end{document}